# Adaptive Technique for Computationally Efficient Time Delay and Magnitude Estimation of Sinusoidal Signals

Santanu Sarma, *Member, IEEE*

*Abstract*— An online, adaptive method of time delay and magnitude estimation for sinusoidal signals is presented. The method is based on an adaptive gradient descent algorithm that directly determines the time delay and magnitudes of two noisy sinusoidal signals. The new estimator uses a novel quadrature carrier generator to produce the carriers for an adaptive quadrature phase detector, which in turn uses an arc tan function to compute the time delay. The proposed method is quite robust and can adapt to significant variation in input signal characteristics like magnitude and frequency imposing no requirement on the magnitudes of the two signals. It even works effectively when the signals have time-varying magnitudes. The convergence analysis of the proposed technique shows that estimate converges exponentially fast to their nominal values. In addition, if the technique is implemented in the continuous time domain, the delay estimation accuracy will not be constrained by the sampling frequency as observed in some of the classical techniques. Extensive simulations show that the proposed method provides very accurate estimates of the time delay comparable to that of the popular methods like Sinc-based estimator, Lagrange estimator, and the Quadrature estimator, as well the magnitude estimate of the input signals at lower signal to noise ratio at appreciably reduced computational cost.

## I. INTRODUCTION

TIME delay estimation (TDE) refers to the methods of estimation of the delay or propagation delay between two (or more) noisy signals received from spatially separated receivers, antennas or sensors. It has attracted much attention in the literature [1]-[2] and arises in many areas including fields as diverse as sonar, radar, biomedicine, geophysics, and ultrasonics. It is widely used in passive sonar where the bearing of a moving target can be determined from the time delay measurements by triangulation. In active sonar, a signal is transmitted into some medium and from the reflected echo information is extracted from the reflecting target. Often, the parameters of interest are the range (location) and velocity of the target, estimates of which may be determined from the differential time delay (DTD) and differential frequency offset (DFO), also termed Doppler shift, between the transmitted and received signals.

The basic idea of time delay estimation can be described as a method to determine the delay $\phi(t)$ between two signals $s(t)$ and its delayed version $s(t+\phi(t))$ in the presence of measurement noise $n(t)$ from the measured signals given by

$$x_1(t) = s(t) + n_1(t), x_2(t) = s(t-\phi(t)) + n_2(t), \quad (1)$$

where $n_1(t), n_2(t)$ are stationary zero-mean white Gaussian noises, $\phi(t)$ is the phase delay between the two signals and can be converted to time delay for known frequency. The objective is thus to estimate the delay $\phi(t)$ from the measured signals $x_1(t), x_2(t)$.

Many off-line and online methods have been proposed for time delay estimation in the past two decades [3-14]. Off-line time domain techniques, such as the Hilbert transform correlation [3], estimators based upon a combination of cross-correlation and auto-correlation [4], and Interpolation [5] and Spline-based techniques [6] are reported where data can be processed later to estimate the time delay. On-line techniques are used when the delay estimate is required to be determined at each new input sample. On the other hand, adaptive on-line techniques [7-13] are usually preferred when the delay is time varying. Generally, the delay estimate is achieved by minimizing the energy of the difference between the measured signals by adaptively choosing the phase delay such that some objective error function is minimized. There are many algorithms for minimizing the objective function such as gradient descent algorithm that has been shown to provide a subsample delay estimate [11]. However, these minimization methods can be computationally expensive and the convergence time may be large [16]. There is also the problem of significant bias in delay estimate when fractional delay filters [9, 10] are used to obtain subsample delay estimate [11] with time varying signal magnitude. With marginal increase in computational cost, methods to reduce the effects of signal magnitude variations and estimation bias are also reported in [12, 13] using a windowed discrete-time quadrature technique and modified DFT technique. Indeed, techniques that reduce the effect of the time varying magnitude and bias, with reduced computational cost and improved convergence properties, are highly desirable.

In this paper, a computationally inexpensive direct, online method is proposed to estimate the time-delay along with joint estimation of the signal magnitudes for sinusoidal signals as desired usually over a continuous time domain in application like medical ultrasound, radars, and sonar. The method is based on a gradient descent adaptive algorithm. It consists of an adaptive quadrature phase detector, a novel quadrature carrier generator, and computation of an arctan function. The adaptive quadrature phase detector provides an estimate of the phasor representation of the delayed signal amplitude with respect to phase delay between the two sinusoids. The tangent of the in-phase and quadrature phase component of this phasor is used to directly compute time delay estimate. On the other hand, the modulus of the quadrature components from the quadrature carrier generator and the phase detector give the estimates of the magnitudes.

## II. PROPOSED TIME DELAY AND MAGNITUDE ESTIMATION SCHEME

The proposed online method of estimating the time delay and the magnitude is show in Fig.1. In this scheme, the quadrature carriers required by the adaptive quadrature phase detector are generated using a magnitude tracking loop as shown in Fig. 2. The normalized carriers are then used in the adaptive quadrature phase detector that synthesizes the delayed signal magnitude as a phasor of the phase delay between the two signals. The in-phase and quadrature phase components of this phasor is used to compute the time delay by computing the arc tan of division of the two signals. The magnitude estimate of both the input signals are obtained by finding the modulus from its sine and cosine components respectively from the quadrature carrier generation and synchronous amplitude demodulation stage as illustrated in Fig. 2 and Fig. 3. Further, a simple look-up table based arc tan computation can be used for finding directly the phase delay between the two signals.

The proposed method differs with the classical estimation techniques in number of ways. Firstly, it provides an online estimate of the time delay formulated in the continuous time domain and the delay estimation accuracy is not constrained by the sampling frequency as in some classical discrete time methods. Moreover, the equivalent discrete time representation would eventually reach the limiting case of continuous time domain for infinite sampling and its implementation using digital logic is also attractive. Secondly, unlike the adaptive methods in [6-14], the estimation of the magnitudes of both the sinusoidal signals is available that can provide useful information, for instance, the signal energy. Thirdly, the proposed method is quite robust and can adapt to significant variation in input signal characteristics like magnitude and frequency. In fact, there is no stringent requirement on the magnitudes of the two signals. Fourthly, the computational cost of the proposed technique is appreciably low and provides comparable performance with popular adaptive techniques.

In this paper the signals $x_1(t), x_2(t)$ are assumed to be real deterministic signal, specifically pure sinusoids that commonly occur in radar, sonar and digital communication applications. These measured signals are expressed as

$$x_1(t) = m_1(t)\cos(\alpha_1(t)) + n_1(t),$$
$$x_2(t) = m_2(t)\cos(\alpha_2(t)) + n_2(t), \quad (2)$$

where $m_1, m_2$ are the magnitude of the signals $s(t), s(t+\phi(t))$ respectively and

$$\alpha_1(t) = \omega_1 t, \alpha_2(t) = \omega_1 t + \phi(t). \quad (3)$$

Generally, in most of the practical cases $\phi(t)$ may be a constant or time varying phase angle. If $\phi(t)$ varies at a rate $\nu$ such that

$$\phi(t) = \nu t + \theta \quad (4)$$

than the phase $\alpha_2(t)$ can be expressed as

$$\alpha_2(t) = \omega_2 t + \theta, \quad (5)$$

where $\omega_2 = \omega_1 + \nu$, and $\nu \ll \omega_1$. The problem consider in this paper is to estimate time delay $\phi(t)$, and the magnitudes $m_1, m_2$ for a known $\omega_1$. The assumption of known frequency pertains to applications like a moving source that emits a constant tone in radar and certain types of underwater acoustic systems.

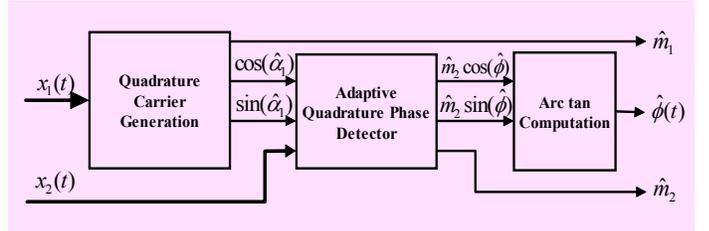

Figure 1: Proposed Time Delay and Magnitude Estimator.

### A. Quadrature carrier generation and magnitude estimation

The purpose of this novel block is to generate the sine and cosine carriers that are of the same magnitude, phase, and frequency as that of input signal $x_1(t)$. This system basically is a magnitude tracking loop for a sinusoidal signal of known frequency. It has an internal oscillator loop that oscillates at the frequency $\omega_1$ of the input signal $x_1(t)$. To achieve this, the gains of the integrators in the oscillator loop are set to $\omega_1$.

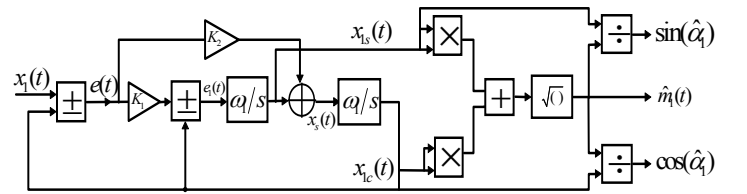

Figure 2: Quadrature carrier generation and magnitude estimation.

A simple proportional control feedback loop is used to control the oscillation amplitude and the quadrature carriers. The dynamics of the loop is expressed by the following equations:

$$\dot{x}_{1s}(t) = -(K_1+1)\omega_1 x_{1c}(t) + \omega_1 K_1 x_1(t),$$
$$\dot{x}_{1c}(t) = \omega_1 x_{1s}(t) - \omega_1 K_2 x_{1c}(t) + \omega_1 K_2 x_1(t). \quad (6)$$

In matrix form of $\dot{X} = AX + BU$, where $X = [x_{1s}, x_{1c}]^T$, $U = x_1$, the matrix $A$ and $B$ are given by

$$A = \begin{bmatrix} 0 & -(K_1+1)\omega_1 \\ \omega_1 & -\omega_1 K_2 \end{bmatrix}, B = \begin{bmatrix} \omega_1 K_1 \\ \omega_1 K_2 \end{bmatrix}. \quad (7)$$

The transfer functions of the system are given by

$$TF_{1s} = \frac{X_{1s}(s)}{X_1(s)} = \frac{\omega_1(K_1 s - \omega_1 K_2)}{(s^2 + \omega_1 K_2 s + (K_1+1)\omega_1^2)},$$
$$TF_{1c} = \frac{X_{1c}(s)}{X_1(s)} = \frac{\omega_1(K_2 s + \omega_1 K_1)}{(s^2 + \omega_1 K_2 s + (K_1+1)\omega_1^2)}. \quad (8)$$

The closed form solution of $x_{1s}, x_{1c}$ for the sinusoidal stimulus $x_1(t) = m_1 \cos(\omega_1 t)$ and initial conditions $x_{1c}(0) = 1, x_{1s}(0) = 0$, are given by

$$x_{1c}(t) = m_1 \cos(\omega_1 t) - \frac{A_1}{\sqrt{K}} e^{-B_1 t} - \frac{A_2}{\sqrt{K}} e^{-B_2 t}$$
$$x_{1s}(t) = -m_1 \sin(\omega_1 t) - \frac{K_2 C_1}{\sqrt{K}} e^{-B_1 t} - \frac{K_2 C_2}{\sqrt{K}} e^{-B_2 t} \quad (9)$$
$$- C_1 e^{-B_1 t} + C_2 e^{-B_2 t},$$

where, the constants $A_1, A_2, C_1, C_2, B_1, B_2$, and $K$ can be easily calculated. For all positive values of $B_1, B_2$ the transients components in (9) die out, leaving the quadrature carriers at steady state. The amplitude of the signal $x_1(t)$ can therefore be obtained by finding the modulus of squares of $x_{1s}$ and $x_{1c}$ as

$$\hat{m}_1 = \sqrt{x_{1s}^2 + x_{1c}^2}. \quad (10)$$

The gains $K_1$ and $K_2$ can be selected such that the transients die out fast and also suppresses the effect of the noise in the generated quadrature carriers. It can be observed from the transfer function in (8) that the gain $K_2$ controls the bandwidth whereas $K_1$ controls the resonance frequency. Moreover, when $K_1 = K_2$ both the transfer functions will have the same magnitude response making the magnitudes of both the carriers equal. For small values of the gains, the band pass transfer function in (8) will have a sharp peak at the resonance frequency, which not only will suppress the noise but also generate the quadrature carriers with equal magnitudes. It is interesting to note that this system produces the Hilbert transform [3] equivalent of the input signal that is used in [12] to produce the phase delay information using fractional delay filter.

*B. Adaptive quadrature phase detector*

The estimation of the time delay is performed mainly by obtaining the phasor representation of the delayed signal's magnitude with respect to the phase delay between the two signals. The proposed adaptive phase detector as shown in Fig. 3 is a closed loop feedback device when compared to the open loop quadrature phase detector as described in [11]. The basic idea of this approach can be understood from the expression

$$x_2(t) = m_2 \cos(\omega_1 t \pm \phi(t))$$
$$= m_2[\cos(\phi(t))\cos(\omega_1 t) \mp \quad (11)$$
$$\sin(\phi(t))\sin(\omega_1 t)].$$

This expression can be interpreted as summation of two amplitude modulated signal with carrier $\cos(\omega_1 t)$ for the first part and $\sin(\omega_1 t)$ for the next. Using two synchronous demodulator [15] with appropriate carrier as in Fig. 3, the estimates of the information signals $m_I = m_2 \cos(\phi(t))$ and $m_Q = m_2 \sin(\phi(t))$ can be derived. The sine and cosine carriers generated by the quadrature carrier generator are used in the adaptive quadrature phase detector to derive estimates of the magnitudes as

$$\hat{m}_I = \hat{m}_2 \cos(\hat{\phi}(t)), \hat{m}_Q = \hat{m}_2 \sin(\hat{\phi}(t)). \quad (12)$$

The individual parts of (11) obtained from the two demodulators are added to produce the estimate of input signal $\hat{x}_2(t)$. The error $e(t)$ between this generated signal and the input signal is minimized in closed loop to make both the loops converge. It can be shown that when $\hat{x}_2(t)$ converges to $x_2(t)$, it produces a phasor representation of the magnitude $m_2$. The modulus of this phasor gives the magnitude of $x_2(t)$.

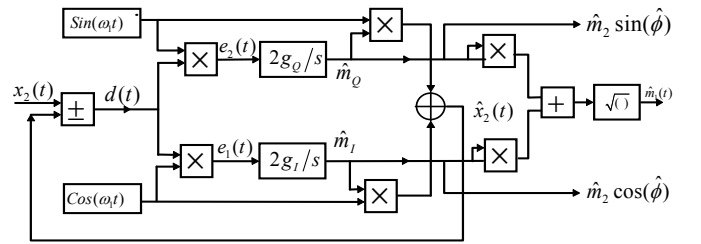

Figure 3: Adaptive quadrature phase detector.

The phasor component of the delayed signal magnitude is multiplied with the respective carriers to reconstruct the input signal. The reconstructed signal for the input and the difference signal between them are given by,

$$\hat{x}_2(t) = \hat{m}(t)\cos(\omega_1 t + \hat{\phi}(t)),$$
$$= \hat{m}_Q(t)\sin(\omega_1 t) - \hat{m}_I(t)\cos(\omega_1 t), \quad (13)$$

$$e(t) = x_2(t) - \hat{x}_2(t) = \tilde{m}_I \cos(\omega_1 t) + \tilde{m}_Q \sin(\omega_1 t), \quad (14)$$

where,

$$\tilde{m}_I = m_I - \hat{m}_I, \tilde{m}_Q = m_Q - \hat{m}_Q. \quad (15)$$

The difference signal is multiplied with sinusoidal carriers as well with a gain $g_Q, g_I$ and integrated respectively to obtain in-phase and quadrature phase magnitude as

$$\dot{\hat{m}}_Q(t) = 2g_Q e(t)\sin(\omega_1 t),$$
$$\dot{\hat{m}}_I(t) = 2g_I e(t)\cos(\omega_1 t). \quad (16)$$

Further, the adaptive quadrature phase detector can be represented as a gradient descent algorithm as

$$\begin{bmatrix} \dot{\hat{m}}_Q(t) \\ \dot{\hat{m}}_I(t) \end{bmatrix} = \begin{bmatrix} -g_Q & 0 \\ 0 & -g_I \end{bmatrix} \nabla e^2(t). \quad (17)$$

where $\nabla e^2(t)$ is the gradient of the squared error.

*C. Stability and Convergence Analysis*

To prove the global convergence of the proposed adaptive quadrature phase detector, we develop the error dynamics of the system and show using Lyapunov direct method that the origin of the error dynamic system is globally asymptotically stable. In other words, this also means that the adaptive quadrature phase detector is globally convergent. The error dynamic equations of (16) can be obtained by differentiating (15) such that

$$\begin{bmatrix} \dot{\tilde{m}}_Q(t) \\ \dot{\tilde{m}}_I(t) \end{bmatrix} = \begin{bmatrix} -g_Q \sin(2\omega_1 t) & -2g_Q \sin^2(\omega_1 t) \\ -2g_I \cos^2(\omega_1 t) & -g_I \sin(2\omega_1 t) \end{bmatrix} \begin{bmatrix} \tilde{m}_Q \\ \tilde{m}_I \end{bmatrix}. \quad (18)$$

We define a Lyapunov function

$$V(\tilde{m}_Q, \tilde{m}_I) = \tilde{m}_I^2/2 + \tilde{m}_Q^2/2, \quad (19)$$

that is positive definite, decrescent, and radially unbounded. The derivative of the Lyapunov function (19) is

$$\dot{V}(\tilde{m}_Q, \tilde{m}_I) = \tilde{m}_I \dot{\tilde{m}}_I + \tilde{m}_Q \dot{\tilde{m}}_Q$$
$$= -2g_I \tilde{m}_I e(t) \cos \omega_1 t - 2g_Q \tilde{m}_Q e(t) \sin \omega_1 t \quad (20)$$
$$= -2ge^2(t), \text{ where } g_I = g_Q = g.$$

Thus, derivative of the Lyapunov function is negative definite i.e. $\dot{V}(\tilde{m}_Q, \tilde{m}_I) < 0$ for

$$g_I = g_Q = g > 0. \quad (21)$$

The equilibrium point of (18) $\tilde{m}_{Qe} = 0, \tilde{m}_{Ie} = 0$ are globally asymptotically stable. This results in infinite equilibrium points of (16), which is obtained by setting right hand side of (16) to zero as

$$m_Q = \hat{m}_Q(t), m_I = \hat{m}_I(t). \quad (22)$$

To prove the convergence of the parameters to their nominal values for the proposed adaptive quadrature phase detector, we represent the estimates as a generalized linear parameter model using a strictly positive real (SPR) error equation [16]

$$e = x_2 - \hat{x}_2 = \hat{M}\theta^{*T}(t)w(t) - \hat{M}\theta^T(t)w(t), \quad (23)$$

where $\theta^* = [m_I \ m_Q]^T$ represents the nominal values of the parameters, $\theta = [\hat{m}_I \ \hat{m}_Q]^T$ corresponds to the parameter estimates, $\hat{M}$ is a SPR transfer function [16, pp 82], and $w(t) = [\sin(\omega_1 t) \ \cos(\omega_1 t)]$ is the input excitation to the system. We use theorem below that states the exponential parameter convergence of the proposed algorithm represented as $\dot{\theta} = -gew$ with SPR error equations (23) for persistence excitations (PE) [16, pp72].

*Theorem:* Let $w: \mathbb{R}_+ \to \mathbb{R}^{2n}$ be piecewise continuous and $[A \ b \ c^T]$ be a minimal realization of a strictly proper, stable, rational transfer function $\hat{M}(s)$, which in other words satisfy the following conditions:

a) $\hat{M}(s)$ is SPR
b) *There exists symmetric positive definite matrices $P$, $Q$, such that $PA + A^T P = -Q, Pb = c$,*

then the gradient algorithm $\dot{\theta} = -gew$ is exponentially stable if $w, \dot{w} \in L_\infty$ and is persistently exciting i.e. there exists $\lambda_1, \lambda_2, \delta > 0$ such that

$$\lambda_2 I \geq \int_{t_0}^{t_0+\delta} w(\tau) w^T(\tau) d\tau \geq \lambda_1 I \quad \forall t_0 \geq 0. \quad (24)$$

where $I$ is an unity matrix.
*Proof:* Refer [16, pp72-87].

We check the PE property of $w$ using (24). We have,

$$PE = \int_{t_0}^{t_0+\delta} w(\tau) w^T(\tau) d\tau,$$

$$PE = \begin{bmatrix} \dfrac{\delta}{2} - \dfrac{\sin 2\omega_1(t_0+\delta)}{4\omega_1} & -\dfrac{\cos 2\omega_1(t_0+\delta) - \cos 2\omega_1 t_0}{4\omega_1} \\ -\dfrac{\cos 2\omega_1(t_0+\delta) - \cos 2\omega_1 t_0}{4\omega_1} & \dfrac{\delta}{2} + \dfrac{\sin 2\omega_1(t_0+\delta)}{4\omega_1} \end{bmatrix} \quad (25)$$

and for $\delta = 2\pi/\omega_1$ we have, $PE = \begin{bmatrix} \pi/\omega_1 & 0 \\ 0 & \pi/\omega_1 \end{bmatrix}$.

Hence, the persistent excitation condition in (24) is satisfied with $\delta = 2\pi/\omega_1$, $0 < \lambda_1 \leq \pi/\omega_1$, $\lambda_2 \geq \pi/\omega_1$. Thus, $w$ is PE, which implies that $\hat{m}_I, \hat{m}_Q$ converge to $m_I, m_Q$ exponentially fast. Using (12), the estimates $\hat{m}_2, \hat{\phi}$ are computed as follows:

$$\hat{m}_2 = \sqrt{(\hat{m}_Q)^2 + (\hat{m}_I)^2}, \hat{\phi} = \tan^{-1}(\dfrac{\hat{m}_Q}{\hat{m}_I}). \quad (26)$$

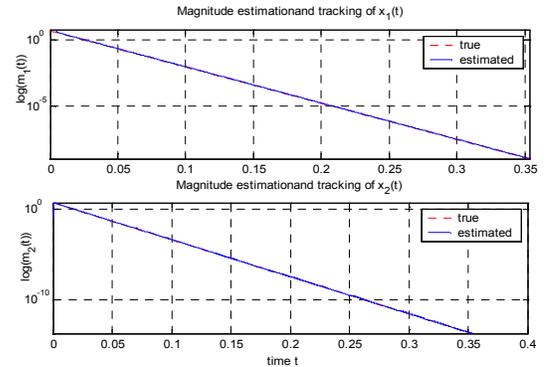

Figure 4: Damped input signals magnitude estimates with damping factors 62.8319 and 94.2478 radian.

Moreover, as $\hat{m}_I, \hat{m}_Q \to m_I, m_Q$ exponentially fast, it follows from (26) that $\hat{m}_2, \hat{\phi}$ converge to $m_2, \phi$ exponentially fast. The rate of convergence of the parameters to their nominal values is related to the value of the adaptation gain given by [16]

$$\alpha = \dfrac{1}{2\delta} \ln \left[ 1 / \left[ 1 - \dfrac{2g\lambda_1}{(1+\sqrt{2n}g\lambda_2)^2} \right] \right], \quad (27)$$

where $\lambda_1, \lambda_2,$ and $\delta$ come from the PE definition (24) and $n$

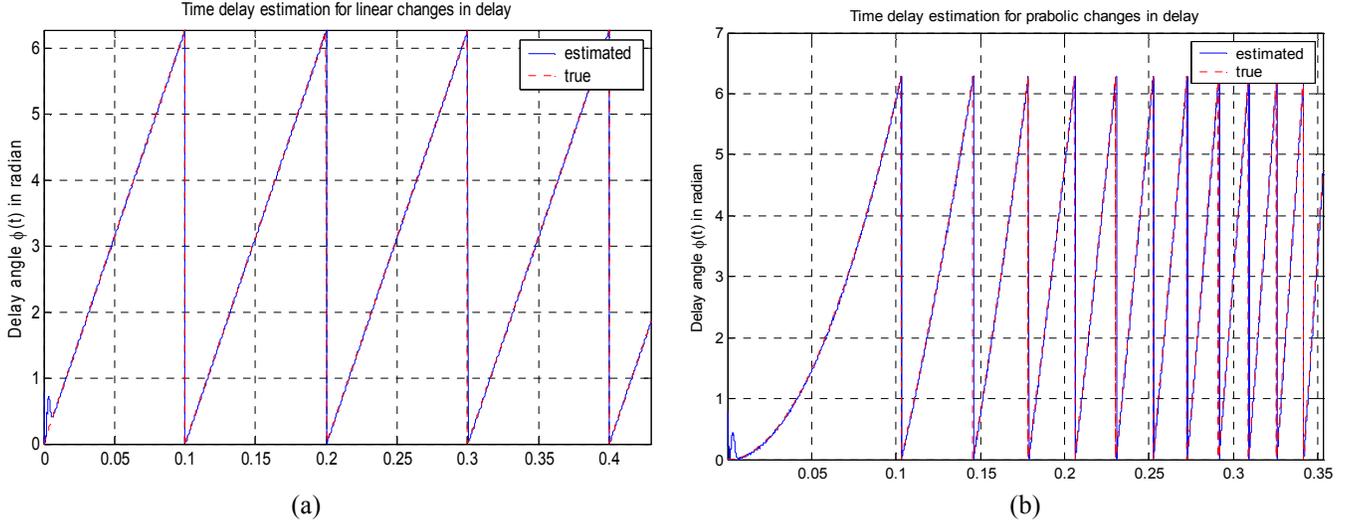

Figure 5: Phase delay estimation without noise for (a) linearly varying delay of 10 Hz Doppler shift with constant magnitudes (b) linearly time varying Doppler at 15 Hz/sec with damped sinusoids with damping factors 62.8319 and 94.2478 radian. X axis shows delay in radian.

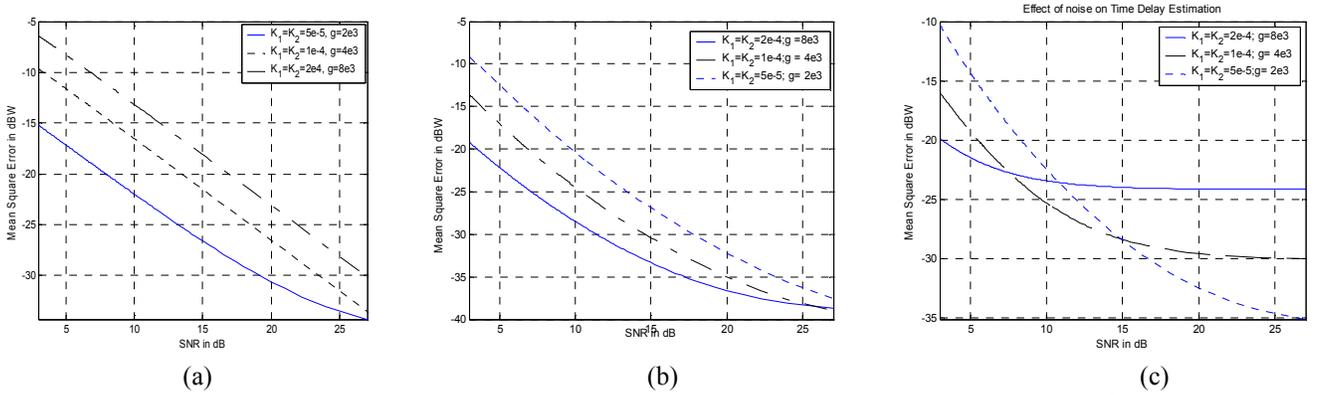

Figure 6: Effect of noise on the estimates of (a) $\hat{m}_1$ (b) $\hat{m}_2$, and (c) time delay $\hat{\phi}$.

is the order of the plant. The relationship in (27) provides the influence of the design parameters on the performance of the proposed algorithm. The constants $\lambda_1, \lambda_2$, and $\delta$ depends on the reference input $m_2$ and the plant being identified.

### III. NUMERICAL SIMULATION AND PERFORMANCE EVALUATION

Extensive numerical simulations had been conducted to evaluate the proposed combined time delay and magnitude estimation scheme in presence of additive white Gaussian noise (AWGN) in both the signals. The performance of the estimator for varied attenuation and time delays are investigated. Studies are also conducted to show the effect of the design parameter on the estimated error. The frequency $\omega_1$ of the source signal $x_1(t)$ is set to 1000 Hz for all simulations.

To elaborate the estimation and tracking performance of the proposed estimator in estimating the magnitude of both the signals $x_1(t)$ and $x_2(t)$, the magnitudes $m_1$ and $m_2$ are considered for damped sinusoidal signals. Figure 3 illustrates the estimation of the magnitude when the signals $x_1(t), x_2(t)$ are considerably damped. The damping factors are 62.8319 and 94.2478 radian respectively for $x_1(t), x_2(t)$. It is evident from Fig. 4 that the proposed estimator has a good tracking capability for time varying amplitudes.

The estimation of the time delay is shown in Fig. 5 for four different cases particularly when the time delay as well as the magnitude is time-varying for the noise free case. It can be observed from the plots in Fig. 4 that the estimator gives a faster tracking of the time delay with linear as well as parabolic variations. It is interesting to note the case of time-varying Doppler shift for the damped sinusoid case. Even when the signal has almost died, the plot on Fig. 5(b) shows that estimation of the time delay is considerably accurate for infinite SNR. This illustrates that the proposed technique can be effective solution when the signals are time varying, due to multipath or similar effects, without needing to resort to any gain control techniques.

The performance of the proposed technique in presence of noise and parameter variations is shown in Fig. 6 for different signal-to-noise-ratio (SNR). The sinusoids amplitude is set to $\sqrt{2}$ and the variance of the noise signal is properly scaled. In these simulations, the time delay is linearly varied resulting in a Doppler shift of 10 Hz between the two signals. The mean square estimation error with variation in SNR for the magnitude $m_1$ is shown in Fig. 6(a). It can be observed from the Fig.6(a) that the noise performance in the estimation of $m_1$ improves with reduction in gains $K_1, K_2$ and SNR. The performance of the magnitude estimation of $x_2(t)$ in presence of additive white Gaussian noise (AWGN) is shown in Fig. 6 (b). Although the exact analytical relation is difficult to find, it is observed that the mean square error in the magnitude estimate $\hat{m}_2$ decreases somewhat linearly with increasing gain and SNR.

The effect of the SNR with parameter variations on time delay estimation is illustrated in Fig. 6 (c). It is observed from Fig. 6(c) that the MSE performance improves with increase gains at low SNR. On the other hand, a reduced set of gains shows better performance at higher SNR. Further, a qualitative evaluation of the performance of the proposed method can be fairly understood by comparing it with some of the popular discrete methods as continuous time domain methods are rare. The root-mean-square (RMS) error versus SNR for the proposed method is compared with that of the Sinc-based estimator [8], Lagrange-based gradient estimator [9], and the adaptive quadrature delay estimator [12] is plotted. The RMS errors were computed from the simulation samples after the convergence of all the methods for similar conditions as in [12]. The gains were set to $K_1 = K_2 = 1e-5$ and $g_I = g_Q = 400$. The adaptation gain of the other techniques is set as in [12] to give a similar rate of convergence. From Fig. 7, it is observed that at low SNR below 10 dB, the proposed adaptive estimator performs relatively better than the other estimators. The results also show that the proposed delay estimator is superior to the Lagrange and Sinc-based estimator across the complete range of SNR.

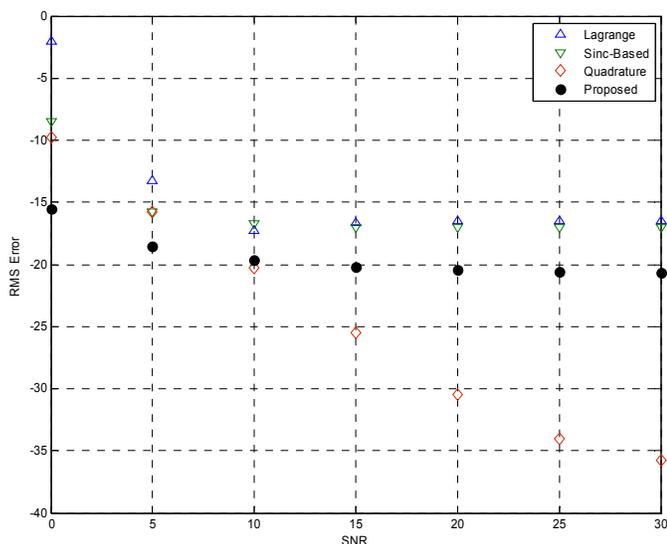

Figure 7: Comparison of the RMS error versus SNR for varying and unequal input amplitudes.

## IV. COMPARISON OF COMPUTATIONAL REQUIREMENTS

One of the appealing features of the proposed estimator is its low computational complexity. The proposed method requires only four integrators, 6 multiplications, 2 divisions and 4 additions. The computations of the magnitudes require two modulus operations, whereas the arc tan computation can be easily realized using look-up tables. This is appreciably less than the popular methods as in [8], [9], [11], and [12]. These schemes use a Lagrange fraction delay filter (FDF) of order 10 in [11] and order3 in [8, 12], which is one order higher than the quadrature carrier generator and takes more computations than the proposed method. Further, at each input sample, Quadrature estimator scheme in [12] also requires the computation of the Hilbert transform, calculation of two scalar products, the two moving averages filters of the quadrature phase detector, and the hard limited inverse tangent along with a division. Moreover, the unbiased estimate of the time delay using [8], [9], and [11] for attenuated amplitudes is not assured [12] and the estimates of the signal magnitude are also not directly available.

## V. CONCLUSION

In this paper a novel, direct, online method of time delay estimation along with combined estimation of the signal magnitude is proposed. The new estimator uses a novel quadrature carrier generator to produce the carriers for an adaptive gradient descent quadrature phase detector, which in turn uses an arc tan function to compute the time delay. The proposed delay estimator offers several advantages over the existing methods, for instance, it provides an online direct estimate of both the time delay and signal magnitudes, rather than an error signal as in other classical adaptive estimators. Moreover, the proposed method works very effectively when the signals have different or time-varying magnitudes and time delays. Further, the delay estimation accuracy is not constrained by the sampling frequency as in some of the classical and subsample adaptive techniques if implemented in continuous time domain. Besides, the computational requirement for the proposed estimator is very attractive compared to the other methods when implemented digitally. As a result of this, the constraints on real-time implementations are relaxed. The convergence analysis of the proposed technique shows that the estimate converges exponentially fast to their nominal values. Extensive numerical simulations were carried out to show the effectiveness of the proposed estimator for static and time-varying magnitudes and delays with and without additive Gaussian noise. It was seen from the simulation results that the proposed method provides very accurate estimates of the time delay comparable to the popular methods like Sinc-based estimator, Lagrange estimator, and the Quadrature estimator and gives relatively better results at low SNR below 10 dB.

## REFERENCES

[1] "Special issue on time delay estimation," IEEE Trans. Acoust. Speech, Signal Processing, vol. 29, no. 3, June 1981.

[2] G.C. Carter, Coherence and Time Delay Estimation: An Applied Tutorial for Research, Development, Test, and Evaluation Engineers, IEEE Press, 1993.

[3] Grennberg and M. Sandell, "Estimation of subsample time delay differences in narrowband ultrasonic echoes using the Hilbert transform correlation," *IEEE Trans. Ultrason., Ferroelect., Freq. Contr.*, vol. 41, pp. 588-595, Sept., 1994.

[4] A.K. Nandi, "On the subsample time delay estimation of narrowband ultrasonic echoes", *IEEE Trans. Ultrason., Ferroelect., Freq. Contr.*, vol.42, pp. 993-1001, Nov, 1995.

[5] Jacob Benesty, , Jingdong Chen, and Yiteng Huang, "Time-Delay Estimation via Linear Interpolation and Cross Correlation," IEEE Transactions on Speech and Audio Processing, Vol. 12, No. 5,pp.509-519, September 2004.

[6] Francesco Viola and William F. Walker "A Spline-Based Algorithm for Continuous Time-Delay Estimation Using Sampled Data" *IEEE transactions on ultrasonics, ferroelectrics, and frequency control*, vol. 52, no. 1, January 2005.

[7] H.H.Chiang and C.L.Nikias, "A new method for adaptive time delay estimation for non-Gaussian signals," *IEEE Trans. Acoust., Speech, Signal Processing*, Vol. 38, no.2, pp.20 9-219. Feb 1990.

[8] H.C. So, P.C. Ching and Y.T. Chan, "A new algorithm for explicit adaptation of time delay," *IEEE Trans. Signal Processing*, vol. 42, pp. 1816-1820, July, 1994.

[9] S.R. Dooley, A.K. Nandi, "Adaptive Subsample Time Delay Estimation Using Lagrange Interpolators," *IEEE Signal Processing Letters*, Vol.6, No. 3, pp.65-67, March 1999.

[10] H. C. So, "Analysis of an Adaptive Algorithm for Unbiased Multipath Time Delay Estimation," *IEEE Transactions on Aerospace and Electronic Systems*, Vol. 39, No. 3, Pp.777-785, July 2003.

[11] D. L. Maskell and G. S. Woods, "The discrete-time quadrature subsample estimation of delay," IEEE Trans. Instrum. Meas., vol. 51, no. 1, pp. 133–137, 2002.

[12] Douglas L. Maskell, *and* Graham S. Woods, *"*Adaptive Subsample Delay Estimation Using a Modified Quadrature Phase Detector," *IEEE Transactions On Circuits And Systems—Ii: Express Briefs*, Vol. 52, No. 10, pp. October 2005.

[13] Douglas L. Maskell, *and* Graham S. Woods "Adaptive Subsample Delay Estimation Using Windowed Correlation," *IEEE Transactions On Circuits And Systems—Ii: Express Briefs*, Vol. 53, No. 6, pp. June 2006.

[14] H.C.So, "A comparative study of two discrete-time phase delay estimators," IEEE Transactions on Instrumentation and Measurement, vol.54, no.6, pp.2501-2504, December 2005.

[15] F.F. Yassa and S.E.Noujaim, "Adaptive Synchronous Amplitude Demodulation," ACSSC,, pp. 107- 111, Mapple Press, 1988.

[16] S.Sastry and M.Bodson, Adaptive Control stability, convergence, and robustness, Prentice Hall, Englewood Cliff, NJ, 1989.